\documentclass[a4paper,11pt]{article}
\usepackage{jheppub} 

\title{Dressing a black hole with a time-dependent Galileon}

\author[a]{Eugeny Babichev}
\author[a,b]{and Christos Charmousis}

\affiliation[a]{Laboratoire de Physique Th\'eorique (LPT), Univ. Paris-Sud, CNRS UMR 8627, F-91405 Orsay, France}
\affiliation[b]{Laboratoire de Math\'ematiques et Physique Th\'eorique (LMPT), CNRS UMR 6083, Universit\'e Francois Rabelais-Tours, France}

\emailAdd{eugeny.babichev@th.u-psud.fr}
\emailAdd{christos.charmousis@th.u-psud.fr}

\abstract{
We present a class of exact scalar-tensor black holes for a shift-symmetric part of the Horndeski action. The action includes a higher order scalar tensor interaction term. We find that for a static and spherically symmetric space-time, the scalar field, if time dependent, can be non-trivial and regular thus circumventing in an interesting way no-hair arguments for gallileons. Furthermore, within this class we find a stealth Schwarzschild and a partially self-tuned de-Sitter Schwarzschild  black hole, both exhibiting a non trivial and regular space and time dependent scalar. In the latter solution the bulk vacuum energy is screened from a necessarily smaller geometric effective de Sitter vacuum via an integration constant associated to the time dependent scalar field. 
}

\arxivnumber{1312.3204}

\begin{document}

\maketitle

\section{Introduction}

Scalar-tensor theories present a generic and well-defined classical alternative to Einstein's General Relativity. Furthermore, recent observational data point towards the tantalizing possibility that GR may be modified at large distances. Indeed, in order for FLRW cosmology to be in accord with observations one needs to assume the presence of a very tiny yet non-zero cosmological constant providing the observed acceleration of the Universe. A tiny cosmological constant is the most economic way to fuel late acceleration, its origin and magnitude however remains a complete puzzle for theoretical physics. 
In a combined effort to attack the large cosmological constant problem in the context of scalar tensor theories~\cite{fab4} it was realized that the most general classical effective scalar-tensor theory was that proposed by Horndeski~\cite{Horndeski}. Horndeski constructed his theory by brute force early on in the 70's but the same result was obtained by studying Galileons in a more intuitive manner in \cite{deffayet}. Either way we are now in full knowledge of the general scalar tensor theory (with no more than second order derivatives in the equations of motion) but very little is known about black hole solutions of Horndeski theories \cite{blaise,Rinaldi:2012vy}. Apart from the fact that such solutions are technically difficult to find, generically scalar-tensor theories do not admit black hole solutions where the scalar field is non-trivial and regular. The difficulty can be summarized in the idea that black holes do not have hair; they are bald objects once they reach a stationary phase having expelled or eaten up all matter surrounding them (see e.g.~\cite{Bekenstein:1998aw}). They are characterized by specific charges, electric, magnetic, angular momentum, charges which can be measured by an observer at infinity. In this paper we will use higher order scalar tensor interactions whose relevant complexity will allow us to evade no hair arguments and construct  in a relatively simple manner analytic black hole solutions where the scalar field will seen to be non trivial and regular. 
 
\section{Constructing Galileon black holes}

Let us consider part of the Horndeski action, 
\begin{equation}\label{action}
S = \int d^4x \sqrt{-g}\left[\zeta R - \eta \left(\partial\phi\right)^2 +\beta G^{\mu\nu}\partial_\mu\phi \partial_\nu\phi - 2 \Lambda \right],
\end{equation}
where $R$ is the Einstein-Hilbert term, $G_{\mu\nu}$ is the Einstein tensor, $\phi$ is a scalar field, $\Lambda$ is 
a cosmological constant term, and $\zeta>0$, $\eta$ and $\beta$ are constants\footnote{In the cosmological context this type of action was studied, e.g. in~\cite{Gubitosi:2011sg}.}.
The above action has shift symmetry with respect to the scalar field, $\phi\to \phi +$const. The presence of the third term, which was referred to as John in~\cite{fab4}, 
of higher order in derivatives, presenting an interaction in between curvature and scalar will be essential for our discussion. 

The variation of the action (\ref{action}) with respect to the metric gives,
\begin{equation}\label{eomg}
\begin{aligned}
\zeta G_{\mu\nu} &-\eta \left(\partial_\mu\phi \partial_\nu\phi -\frac12 g_{\mu\nu}(\partial\phi)^2 \right)  +g_{\mu\nu}\Lambda \\
	&+\frac{\beta}2 \left( (\partial\phi)^2G_{\mu\nu} + 2 P_{\mu\alpha\nu\beta} \nabla^\alpha\phi \nabla^\beta\phi \right. \\
	 & \left.   +  g_{\mu\alpha}\delta^{\alpha\rho\sigma}_{\nu\gamma\delta}\nabla^\gamma\nabla_\rho\phi \nabla^\delta\nabla_\sigma\phi \right)
	=0,
\end{aligned}
\end{equation}
where $P_{\alpha\beta\mu\nu}$ is the double dual of the Riemann tensor, 
$
	P_{\alpha\beta\mu\nu} = -\frac14 \epsilon_{\alpha\beta\rho\sigma} R^{\rho\sigma\gamma\delta}\epsilon_{\mu\nu\gamma\delta}.
$
The variation of the action with respect to $\phi$ can be rewritten in the form of a current conservation, as a consequence of the shift symmetry of 
the action,
\begin{equation}\label{eomJ}
 \nabla_\mu J^\mu =0,\;\; J^\mu = \left( \eta g^{\mu\nu} -\beta G^{\mu\nu} \right) \partial_\nu\phi.
\end{equation}
Note that \eqref{eomJ} contains a part of the metric field equations, namely, an Einstein-Hilbert plus cosmological constant term. 
With this definition at hand it is interesting to discuss two regularity conditions.
The first stems from a no-hair argument for Galileons, \cite{Hui:2012qt} where, the static spherically symmetric configurations of certain Galileons with shift invariance were argued to admit a no-hair theorem (see also~\cite{Germani:2011bc}).
The key point in their argument is the physical requirement that the square of the Noether current, 
$J^2 \equiv J_\mu J^\mu$, does not diverge at the horizon. This fixes the only nontrivial component $J^r$, to zero, 
and then by use of translation invariance it can be argued that  $\phi=$const everywhere. 
In our case we also have a conserved current and for $\phi =$const, $J^\mu =0$. 
There is, however, a second option yielding $J^2$ finite at the horizon where by imposing,
\begin{equation}\label{condition} 
\beta G^{rr}-\eta g^{rr}=0,
\end{equation}
for a static configuration. This condition is, rather conveniently as we will see, one of the Einstein plus cosmological constant equations of motion. 
The condition (\ref{condition}) automatically kills the $J^r$ component of the current without implementing a constraint on the scalar field.

The second regularity condition is that the scalar field does not explode at the horizon. 
A typical example where this latter condition is not met is the BBMB black hole \cite{BBMB}. There the black hole geometry is everywhere regular (apart from the central singularity) but the scalar field explodes at the horizon location. 
As we will see below, our first condition is in general not sufficient to satisfy the second.
In fact to meet both requirements we will have to additionally assume that the scalar field is also time-dependent $\phi=\phi(t,r)$, unlike the metric which we will assume to be static and spherically symmetric,
\begin{equation}\label{metric}
	ds^2 = -h(r)dt^2 + \frac{dr^2}{f(r)} + r^2 d\Omega^2.
\end{equation}
Allowing scalar time dependence modifies the no hair argument as now the time component of the Noether current will generically be non-zero, 
\begin{equation}\label{Jt}
 J^t = \left( \eta g^{tt}-\beta G^{tt} \right) \dot\phi(t,r),
\end{equation}
and can potentially source regularity problems for the current.
The aim of this paper is to construct non-trivial black holes that satisfy the above two conditions, i.e. allow for a regular current and 
non trivial regular scalar field in a static and spherically symmetric black hole geometry.   

For the ansatz (\ref{metric}) the $tr$ component of (\ref{eomg}) reads,
\begin{equation}\label{tr}
	\frac{\beta \phi'}{r^2}\left( \frac{rfh'}{h}+\left(f-1-\frac{\eta r^2}{\beta}\right)\dot\phi -2rf\dot\phi'\right)=0,
\end{equation}
where dot $= \partial/\partial t$  and prime $= \partial/\partial r$. 
Apart from the obvious $\phi'=0$, the expression inside the parentheses can be integrated  to give,
\begin{equation}\label{trphi}
 \phi(t,r) = \psi(r) + q_1(t) e^{X(r)},
\end{equation}
where,
\begin{equation}\label{X}
X(r) = \frac12\int dr\left( \frac{1}{r}-\frac{1}{rf}-\frac{\eta r}{\beta f} +\frac{h'}{h}\right).
\end{equation}
and we note that, $\beta G^{rr}-\eta g^{rr}=-2 \beta f^2 X'/r$.
With this result at hand, substituting~(\ref{trphi}) with (\ref{X}) into (\ref{eomJ}),
it is easy to show that $q_1(t)$ satisfies the ODE,
\begin{equation}\label{ddm}
\ddot q_1(t) =C_1 q_1(t)+C_2,
\end{equation}
with $C_1$ and $C_2$ integration constants. 
Setting $q_1(t)=0$ and assuming the trivial configuration for the scalar, $\phi'=0$, the scalar field equation (\ref{eomJ}) is straightforwardly satisfied,
while the Einstein equations~(\ref{eomg}) are satisfied by the Schwarzschild metric.
This is similar to what happens in most scalar-tensor theories and in particular, in Brans-Dicke theory. 

Let us now turn to non-trivial solutions where
we note that, setting~(\ref{condition})
and $q_1(t)= q t$ 
render the scalar field equation, $\partial_t \left(\sqrt{-g}J^t\right)+\partial_r \left(\sqrt{-g}J^r\right)=0$, redundant.
One can also think of this Ansatz as switching off the constant associated to primary scalar hair of $\phi$.
Note that the linear dependence of $\phi(t,r)$ on time ``passes through'' the equations of motion, leaving ODEs rather than the original PDEs due to the shift symmetry of the Lagrangian. 

Under these observations we consider the following subclass of (\ref{trphi}),
\begin{equation}\label{anphi}
 \phi(t,r) = q\, t + \psi(r)
\end{equation}
with~(\ref{condition})  which gives $X(r)=$const, in (\ref{X}).
The same ansatz has been applied for the study of test galileon fields in various physical setups~\cite{Babichev:2010kj}. 
Note that, (\ref{anphi}) and (\ref{condition}) satisfy both the $tr$ component (\ref{tr}) and the scalar field equation. 
Using (\ref{condition}) we get,
\begin{equation}\label{f}
	f = \frac{ (\beta+\eta  r^2) h}{ \beta (rh)'}.
\end{equation} 

Since in general $q\neq 0$, the time component of the Noether current~(\ref{Jt}), is also non-zero, 
and one may worry that $J^2$ is in fact diverging at the horizon.
However using~(\ref{f}) gives,
\begin{equation}\label{Jt2}
	J^t = \frac{\left( 2\eta r h -2 \beta h' - r(\beta+\eta r^2) h''\right)q}{r (rh)'^2}.
\end{equation}
Therefore, at the horizon, where $h=0$, we have $J^2 = g_{tt} J^t J^t=0$ for the ansatz (\ref{anphi}) unless $(rh)'=0$ (as is the case for an extremal black hole). 

The first regularity condition satisfied we can now move on, substituting (\ref{f}) and (\ref{anphi}) into the $rr$ component of 
\eqref{eomg} in terms of $\psi'$, to get,
\begin{equation}\label{psi}
	\psi' = \pm \frac{\sqrt{r}}{h(\beta+\eta r^2)}\left(q^2\beta (\beta+\eta r^2) h'-\frac{\lambda}{2}(h^2r^2)'\right)^{1/2}.
\end{equation}
where we introduced a notation $\lambda\equiv \zeta\eta + \beta\Lambda$. 
Finally, substituting (\ref{anphi}), (\ref{f}) and (\ref{psi}) in the $tt$ component of the Einstein equation (\ref{eomg}), 
one obtains  a non-linear second order ODE on $h(r)$.
Under the substitution,
\begin{equation}\label{h}
	h(r) = -\frac{\mu}{r} +\frac{1}{r}\int \frac{k(r)}{\beta+\eta r^2}dr,
\end{equation}
where $\mu$ is an integration constant \eqref{h} can be further integrated 
to give $k(r)$ as a solution of the third order algebraic equation,
\begin{equation}\label{k}
	q^2\beta (\beta+\eta r^2)^2 - \left(2\zeta\beta+\left( 2\zeta\eta -\lambda \right) r^2\right) k + C_0 k^{3/2} =0,
\end{equation}
where $C_0$ is the second integration constant. 
To sum up the equation (\ref{k}), together with (\ref{h}), (\ref{psi}), (\ref{f}) and (\ref{anphi}) gives a class of solutions for the considered theory satisfying non-trivially the first of the two regularity criteria. 

The regularity of the metric and the scalar field at the horizon can be
conveniently checked by use of the generalized Eddington-Finkelstein coordinates, with the advanced time coordinate, $v$,
\begin{equation}\label{v}
	v = t + \int ( fh )^{-1/2} dr.
\end{equation}
One finds from (\ref{metric}) and (\ref{v}),
\begin{equation}\label{metricEF}
	ds^2 = - h dv^2 +2\sqrt{h/f}\, dv dr + r^2 d\Omega^2.
\end{equation}
For our class of solutions, Eq.~(\ref{f}) is satisfied, 
therefore the metric (\ref{metric}) is regular provided that  $(rh)$ is not zero\footnote{Note that $\beta+\eta  r^2 = 0$ and $h\neq 0$ implies a coordinate 
singularity rather than a physical one. By changing the radial coordinate the metric  can be transformed into a regular form. This can also be checked by direct calculation of curvature invariants.}.

We are now ready to attack regularity of the scalar field on the horizon.
Note that although the radial part of the solution (\ref{psi}) seems to be divergent at the horizon, the same holds for coordinate time $t$. 
Hence given the time dependence the correct way to see the horizon behavior of $\phi$ is to use the regular coordinates $(v,r)$ 
(see e.g.~\cite{Jacobson:1999vr,AyonBeato:2004ig}), in which the metric takes the form (\ref{metricEF}).
We can rewrite the scalar~(\ref{anphi}) with (\ref{psi}), and then, using (\ref{v}), and expanding near the horizon, $h= 0$, we find the value of 
the scalar at the horizon, $\phi_{hor}$,
\begin{equation}\label{phiEF}
\begin{aligned}
	\phi_{hor} &= qv +\text{const}\\
	-& \frac{q}2\sqrt{\frac{\beta r}{h'(\beta+\eta r^2)}}\left( 1 + \frac{\lambda r^2 h'}{q^2\beta(\beta+\eta r^2)} \right)\Big|_{hor},
\end{aligned}
\end{equation}
where the last term is evaluated at the future horizon $h=0$ for the plus sign in \eqref{psi}.
One sees then that for time-dependent solutions, $q\neq 0$, the scalar field is actually regular
(provided also that $h'(\beta+\eta r^2)\neq 0$ at the horizon). 
Implicitly we know analytically all the solutions belonging to this class. 
Let us concentrate on some explicit and simple solutions.

{ \it Stealth Schwarzschild black hole}. We start by a subset of Fab 4 theory \cite{fab4} where $\Lambda=\eta =0$ and the scalar field acquires dynamics via the $G^{\mu\nu}\partial_\mu\phi \partial_\nu\phi$ (John) term. 
In this case (\ref{k}) does not depend on $r$ and $k=$const which is just a gauge choice. Choosing $k=\beta$, we obtain from (\ref{h}) and (\ref{f}),
\begin{equation}\label{fh1}
 f  =h = 1- \frac{\mu}{r},
\end{equation}
and the metric is isometric to a Schwarzschild metric! However, the scalar field is not trivial and also regular. 
Indeed from (\ref{psi}) we obtain, $\psi'=\pm q \sqrt{\mu r}/(r-\mu)$.
Integrating $\psi'$, and taking into account~(\ref{anphi}) gives,
\begin{equation}\label{phi1}
	\phi_{\pm}=qt\pm q\mu\left[2\sqrt{\frac{r}\mu}+ \log\frac{\sqrt{r}-\sqrt{\mu}}{\sqrt{r}+\sqrt{\mu}}\right]+\phi_0
\end{equation}
One can explicitly check that the solution (\ref{phi1}) with the plus sign does not diverge on the future horizon (whereas the solution with the minus sign is regular on the past horizon). Indeed the transformation (\ref{v}) in the case (\ref{fh1}) reads,
$
v=t+r+\mu\log(r/\mu-1),
$
and using (\ref{phi1}) one finds,
\begin{equation}
	\phi_+ = q\left[v-r + 2\sqrt{\mu r} - 2\mu \log\left(\sqrt{\frac{r}{\mu}}+1\right)\right]+\text{const},
\end{equation}
which is manifestly regular at the horizon, $r=\mu$. Hence the scalar field does not back-react on the metric and the current is zero. 
The derivatives of the scalar field which are the relevant terms appearing in the action \eqref{action} are also finite at infinity. 
{\footnote{For flat spacetime, $\mu =0$, $\phi=qt$ the scalar degree of freedom is in fact strongly coupled. 
For non-zero mass $\mu$, however, the above solution avoids this problem.} 
Such a configuration, where the absence of pathologies depends on the background solution, is typical for non-canonical theories, including Galileons.}
We have thus constructed a regular Fab 4 black hole with GR geometry and a regular interacting scalar. 

{\it Schwarzschild black hole in an Einstein static universe.} 
Let us choose the parameters of the Lagrangian so that $\eta\neq 0$ and $\Lambda\neq 0$, with $\lambda=\zeta\eta +\beta\Lambda=0$.
In what follows we do not restrict ourselves to the case $\eta>0$ --- the would-be healthy kinetic term in the absence of the John term. 
As we already mentioned above, the spin-0 degree of freedom also acquires dynamics via the kinetic mixing with the spin-2 graviton, thanks to the John term. 
Therefore the condition for the solution to be ghost-free is different from the standard case.
See e.g. \cite{Deffayet:2010qz}, where a galileon model has been shown to be stable on cosmological solutions, although the standard kinetic term has a 'wrong'  sign.
We also take $C_0=0$ in (\ref{k}) for simplicity (other algebraic solutions are tedious but also easily obtained).
The scalar now backreacts on the geometry. Indeed we find
\begin{equation}\label{fh2}
	h = 1- \frac{\mu}{r}, \;\;  f=\left(1- \frac{\mu}{r}\right)\left(1+\frac{\eta r^2}{\beta}\right),
\end{equation}
whereas the radial part of the scalar field is given by
\begin{equation}\label{psi2}
\psi' = \pm \frac{q}{h}\sqrt{\frac{\mu}{r (1+\frac\eta\beta r^2) }}
\end{equation}
where $q^2=2\zeta/\beta$ is fixed while the current vanishes only at the horizon since $J^t = 2\eta$.
The metric is therefore regular apart from $r=0$. The solution is not asymptotically usual.
In fact in the absence of the central mass, $\mu =0$ and $\beta\eta<0$ the metric~(\ref{fh2})  is the Einstein static universe with 
Gaussian curvature $-\eta/\beta>0$. 
With non-zero  $\mu$ this solution therefore describes a Schwarzschild black hole embedded in a static Einstein universe. 
Provided $\eta<0$ and $\beta>0$ the scalar field is regular everywhere, apart from the 2-sphere at $r=\sqrt{-\beta/\eta}$, 
where $\phi$ has a cusp $\sim (r-\sqrt{-\beta/\eta})^{1/2}$.

On the other hand, for $\eta>0$ and $\beta>0$, the metric~(\ref{fh2}) describes a static universe with a negative constant curvature. 
The metric~(\ref{fh2}) and the scalar field~(\ref{psi2}) are regular everywhere but $r=0$.

{\it Self-tuned Schwarzschild-de-Sitter.} 
It is interesting to seek solutions of de Sitter or anti de Sitter asymptotics~\footnote{We thank Tony Padilla for discussions on this particular topic.}. In order to achieve this we can ask for $h(r)=f(r)$ which in turn using \eqref{h} and \eqref{f} means that $k(r)=\frac{(\beta+\eta r^2)^2}{\beta}$.
This turns out to be solution of (\ref{k}) by taking $q^2 = \lambda/(\beta\eta)$ and $C_0 = (2\zeta\eta-\lambda)\sqrt{\beta}/\eta$. 
The metric coefficients take the following form,
\begin{equation}\label{fh3}
 f  =h = 1- \frac{\mu}{r} + \frac{\eta}{3\beta}r^2,
\end{equation}
which is just a (a)dS-Scharzschild metric for an effective cosmological constant $\Lambda_\text{eff} = -\zeta \eta/\beta$!
The solution for the scalar then becomes,
\begin{equation}\label{psi3}
	\psi' = \pm\frac{q}{h}\sqrt{1-h}.
\end{equation}
Let us take for definiteness $\eta>0$ (standard kinetic term), then the above expression makes sense when $\beta < 0$ and $\lambda<0$. 
The first condition implies positive effective  cosmological constant $\Lambda_\text{eff} = \zeta\eta/|\beta|$ in (\ref{fh3}), 
while a negative $\lambda$ insures that~(\ref{psi3}) is real.

The solution (\ref{fh3}) and (\ref{psi3}) has certain self-tuning properties since it hides from the space-time metric the vacuum cosmological constant $\Lambda$. The value of $q$ is fixed by the bare cosmological constant, $q^2 \eta= \Lambda-\Lambda_{eff}>0$, leaving $\Lambda_{eff}$ a geometrical quantity. However, the effective cosmological constant is given by geometric coupling constants in the action essentially the Einstein Hilbert and John term. In order for $\Lambda_{eff}$ to be small one creates an hierarchy or fine tuning inevitably due to the small value of the observed cosmological constant. 

{\it Static-scalar solution.} 
Here we give a solution which has $J^2 =0$, but at the same time does not satisfy the second condition. 
For this solution $\phi$ is not regular at the horizon.
This is essentially due to the fact that we set $q =0$, so that the solution is time-independent.
Similar to the case of the stealth Schwarzschild solution above, the constant $C_0$ in (\ref{k}) amounts for a gauge choice. 
Taking for convenience $C_0 = \sqrt{\beta (4\zeta^2\eta^2 - \lambda^2)}/\eta$, we obtain, 
\begin{equation}\label{Rinh}
\begin{aligned}
	h & = 
	1-\frac{\mu}{r} + \frac{\eta}{3\beta}\frac{2\zeta\eta -\lambda}{2\zeta\eta + \lambda}r^2 \\
	 & + \frac{\lambda^2}{4\zeta^2\eta^2 - \lambda^2}\frac{\arctan(r\sqrt{\eta/\beta})}{r\sqrt{\eta/\beta}},
\end{aligned}
\end{equation}
and $f$ can be computed from~(\ref{Rinh}) and (\ref{f}). The scalar field is,
\begin{equation}\label{Rinphi}
	\psi'^2  = -\frac{\zeta\eta^3 r^2 \left( 2\zeta\beta+(2\zeta\eta-\lambda) r^2\right)^2}{\beta(4\zeta^2\eta^2-\lambda^2)(\beta+\eta r^2)^3h}.
\end{equation}
This solution is a generalization of the solution found in~\cite{Rinaldi:2012vy} for non-zero $\Lambda$.
By construction, the current does not diverge at the horizon, $J^2 =0$, but the scalar field does (therefore this solution is not in contradiction with~\cite{Germani:2011bc}, where 
the scalar field was assumed to be non-divergent).

\section{Conclusions}
In summary, we have given here a novel class of scalar-tensor black holes that present several surprising properties that to our knowledge have not been encountered before. 
They can have pure GR space-time geometry in the simplest of cases but also a more complex form as solutions of the third order polynomial equation (\ref{k}). This is a subject of further study. 
When the solutions have GR space-time geometry, there is no Vainshtein radius associated with the scalar tensor solution, which in the case of minimal coupling of the scalar field to matter is a general property, since matter feels a pure GR metric. 
The black hole geometry is GR-like everywhere. In fact for the stealth Schwarzschild and for the self-tuning de Sitter it would be impossible to make the difference in between a GR and a scalar tensor black hole, without perturbations and of course assuming that we are in the physical frame and thus matter only couples to the metric. Perturbation theory will inevitably re-emerge the scalar field, giving in principle differing phenomenology in between the two. Furthermore, it will tell us of the stability of these black holes. One may wonder if this GR stealth configuration is the stable one. 
The second property is that the scalar field although non-trivial is regular especially where other scalar tensor black holes fail \cite{BBMB}, at the future horizon. At infinity the scalar itself, diverges but not its derivatives which are the ones present at the action. They are finite at asymptotic infinity. The situation for the Schwarzschild solution we depict here is somewhat similar to the axionic black holes of \cite{Bowick:1988xh} where the only effect of the axion is at asymptotic infinity and the bulk solution is a Schwarzschild black hole. Finally the scalar can be space and time dependent whereas space-time is static. More importantly the time dependence here is linear and given the translational invariance of the action, no time derivatives appear whatsoever. This effect is also similar to axionic solutions with non zero axionic charge found recently in adS space-time \cite{Bardoux:2012aw}. There, the axionic fields have linear time dependence in the flat horizon coordinates and a careful Hamiltonian analysis shows the existence of axionic charge at infinity. If this is the case here for the effect of the constant $q$ then this would be qualified as hair rather as a novel scalar charge measured out at infinity and associated to the shift symmetry of the teary.  It is thus difficult to really qualify these solutions as hairy or non-hairy. As we saw, to evade no hair arguments we switch off the radial scalar charge by imposing a geometric condition that renders the scalar field equation redundant. Time dependence, however, provides a new integration constant $q$ that makes the scalar non-trivial and also regular. We could qualify this as secondary hair since the scalar couples non-trivially to space-time geometry via the $G^{\mu\nu}\partial_\mu\phi \partial_\nu\phi$ (John) term. These and other questions we leave for further study. 

\acknowledgments
We would like to thank Gilles Esposito-Far\`ese, Sandro Fabbri and Tony H Padilla for useful and interesting discussions.



\end{document}